\documentclass[runningheads]{llncs}
\usepackage[utf8]{inputenc}
\usepackage[english]{babel}
\let\proof\relax
 
\usepackage{amsthm}
\usepackage{mathtools}
\usepackage{float}

\usepackage{tikz}
\tikzstyle{vertex}=[circle, draw, inner sep=0pt, minimum size=5pt]
\newcommand{\vertex}{\node[vertex]}
\usetikzlibrary{decorations.markings}
\usetikzlibrary{decorations.pathreplacing}
\usetikzlibrary{arrows.meta}
\usepackage{anyfontsize}
\usepackage{graphicx}
\usepackage{amsmath}
\usepackage{amsfonts}
\usepackage{xcolor}
\usepackage{amssymb}
\usepackage{amsmath}
\usetikzlibrary{positioning}
\newcommand{\boundellipse}[3]
{(#1) ellipse (#2 and #3)
}
\usetikzlibrary{shapes.geometric}
\tikzstyle{square}=[draw, shape=regular polygon, regular polygon sides=4,draw,inner sep=0pt,minimum
size=0.225cm]
\tikzstyle{triangle}=[draw, shape=regular polygon, regular polygon sides=3,draw,inner sep=0pt,minimum
size=0.3cm]
\definecolor{magenta}{rgb}{0.8, 0.0, 0.8}
\definecolor{cyan}{rgb}{0.0, 1.0, 1.0}
\definecolor{green1}{rgb}{0, 1, 0}
\definecolor{green}{rgb}{0, 1, 0}
\definecolor{brown}{rgb}{0.65, 0.16, 0.16}
\definecolor{aquamarine}{rgb}{0.5, 1.0, 0.83}
\definecolor{battleshipgrey}{rgb}{0.52, 0.52, 0.51}
\begin{document}
\title{On Structural Parameterizations of the Offensive  Alliance Problem}
%
%
\author{Ajinkya Gaikwad
 \and Soumen Maity
 }
\authorrunning{A.\,Gaikwad and S.\,Maity}
%
\institute{Indian Institute of Science Education and Research, Pune, India 
\email{\texttt{ajinkya.gaikwad@students.iiserpune.ac.in;}}
\email{\texttt{soumen@iiserpune.ac.in}}
}
\maketitle              
\begin{abstract}  The {\sc Offensive Alliance} problem has been studied extensively during the last twenty years. A set $S\subseteq V$ of vertices is an offensive alliance  in an undirected graph $G=(V,E)$ if
 each  $v\in N(S)$ has at least as many neighbours in $S$ as it has neighbours (including itself) not in $S$.   We study the parameterzied complexity of the 
{\sc Offensive Alliance} problem, where the aim is to find a minimum size
offensive alliance.
Our focus here lies on parameters that measure the structural properties of the 
input instance. We enhance our understanding of the problem from the viewpoint of parameterized complexity by showing that 
the problem is W[1]-hard parameterized by a wide range of 
fairly restrictive structural parameters such as the feedback vertex set number, treewidth, pathwidth, and treedepth of the input graph.

\end{abstract}

\section{Introduction}
In real life, an alliance is a collection of people, groups, or states such that the union is 
stronger than individual. The alliance can be either to achieve some common purpose, to protect against 
attack, or to assert collective will against others. This motivates the definitions of defensive and offensive
alliances in graphs. 
The properties of alliances in graphs were first studied by Kristiansen, Hedetniemi, and Hedetniemi 
\cite{kris}. 
They introduced defensive, offensive and powerful alliances. 
The alliance problems have been studied extensively during last twenty years \cite{frick,SIGARRETA20061345,chel,ROD,SIGA}, and generalizations called $r$-alliances are also studied \cite{SIGARRETA20091687}. 
Throughout this article, $G=(V,E)$ denotes a finite, simple and undirected graph of order $|V|=n$. The subgraph induced by 
$S\subseteq V(G)$ is denoted by $G[S]$. For a vertex $v\in V$, we use $N_G(v)=\{u~:~(u,v)\in E(G)\}$ to denote the (open) neighbourhood 
of vertex $v$ in $G$, and $N_G[v]=N_G(v)\cup \{v\}$ to denote the closed neighbourhood of $v$. The degree $d_G(v)$ of a vertex 
$v\in V(G)$ is $|N_G(v)|$. For a subset $S\subseteq V(G)$, we define its closed neighbourhood as $N_G[S]=\bigcup_{v\in S} N_G[v]$ and its 
open neighbourhood as $N_G(S)=N_G[S]\setminus S$. 
For a non-empty subset $S\subseteq V$ and a vertex $v\in V(G)$, $N_S(v)$ denotes the set of neighbours of $v$ in $S$, that is, 
$N_S(v)=\{ u\in S~:~ (u,v)\in E(G)\}$.  We use $d_S(v)=|N_S(v)|$ to denote the degree of vertex $v$ in $G[S]$. 
The complement of the vertex set $S$ in $V$ is denoted by $S^c$.\\


\noindent A non-empty set 
$S\subseteq V$ is a  {\it defensive  alliance} in $G$ if 
$d_S(v)+1\geq d_{S^c}(v)$ for all $v\in S$.
Since each vertex in a defensive alliance $S$ has at least as many vertices from 
its closed neighbor in $S$ as it has in $S^c$, by strength of numbers, 
we say that every vertex in $S$ can be {\it defended} from possible attack by vertices 
in $S^c$. 

\begin{definition}\rm
A non-empty set 
$S\subseteq V$ is an offensive alliance in $G$ if 
$d_S(v)\geq d_{S^c}(v)+1$ for all $v\in N(S)$.
\end{definition}
\noindent Since each vertex in $N(S)$ has more neighbors in $S$ than in $S^c$, we say that every vertex in $N(S)$ is {\it vulnerable} to possible attack by vertices in $S$. Equivalently, since an attack by the vertices in $S$ on the vertices in $V \setminus S$ can result in no worse than a “tie” for $S$, we say that $S$ can effectively attack $N(S)$.

\begin{definition}\rm
A non-empty set 
$S\subseteq V$ is a strong offensive alliance in $G$ if 
$d_S(v)\geq d_{S^c}(v)+2$ for all $v\in N(S)$.
\end{definition}

 
\noindent In this paper, we consider {\sc Offensive Alliance}, {\sc Exact Offensive Alliance}
and {\sc Strong Offensive Alliance} problems under structural parameters. We define these problems as follows:  
    \vspace{3mm}
    \\
    \fbox
    {\begin{minipage}{33.7em}\label{FFVS0 }
       {\sc  Offensive Alliance}\\
        \noindent{\bf Input:} An undirected graph $G=(V,E)$ and an  integer $k\geq 1$.
    
        \noindent{\bf Question:} Is there an  offensive alliance $S\subseteq V(G)$ such that 
        $1\leq |S|\leq k$?
    \end{minipage} }
    \vspace{3mm}
    \\
    \fbox
    {\begin{minipage}{33.7em}\label{FFVS1 }
       {\sc Exact Offensive Alliance}\\
        \noindent{\bf Input:} An undirected graph $G=(V,E)$ and an  integer $k\geq 1$.
    
        \noindent{\bf Question:} Is there an  offensive alliance $S\subseteq V(G)$ such that 
        $|S|= k$?
    \end{minipage} }
    \vspace{3mm}
    \\
    \fbox
    {\begin{minipage}{33.7em}\label{FFVS2 }
       {\sc Strong Offensive Alliance}\\
        \noindent{\bf Input:} An undirected graph $G=(V,E)$ and an  integer $k\geq 1$.
    
        \noindent{\bf Question:} Is there a strong offensive alliance $S\subseteq V(G)$ such that 
        $1\leq |S|\leq k $?
    \end{minipage} }
    \vspace{3mm}

\noindent For standard notations and definitions in graph theory, we refer to West \cite{west}.
     For the standard concepts in parameterized complexity, see the recent textbook by Cygan et al. \cite{marekcygan}.
     The graph parameters we explicitly use in this paper are vertex cover number, feedback vertex set number, pathwidth, treewidth and 
treedepth. 

\begin{definition} \rm 
 The {\it vertex cover number} is the size of a minimum vertex cover in a graph $G$ and it is denoted by
 $vc(G)$.
\end{definition}

    \begin{definition} {\rm
        For a graph $G = (V,E)$, the parameter {\it feedback vertex set} is the cardinality of a smallest set $S \subseteq V(G)$ such that the graph $G-S$ is a forest and it is denoted by $fvs(G)$.}
    \end{definition}

We now review the concept of a tree decomposition, introduced by Robertson and Seymour in \cite{Neil}.
Treewidth is a  measure of how “tree-like” the graph is.
\begin{definition}\rm \cite{Neil} A {\it tree decomposition} of a graph $G=(V,E)$  is a tree $T$ together with a 
collection of subsets $X_t$ (called bags) of $V$ labeled by the vertices $t$ of $T$ such that 
$\bigcup_{t\in T}X_t=V $ and (1) and (2) below hold:
\begin{enumerate}
			\item For every edge $(u,v) \in E(G)$, there  is some $t$ such that $\{u,v\}\subseteq X_t$.
			\item  (Interpolation Property) If $t$ is a vertex on the unique path in $T$ from $t_1$ to $t_2$, then 
			$X_{t_1}\cap X_{t_2}\subseteq X_t$.
		\end{enumerate}
	\end{definition}
	
	
\begin{definition}\rm \cite{Downey} The {\it width} of a tree decomposition is
the maximum value of $|X_t|-1 $ taken over all the vertices $t$ of the tree $T$ of the decomposition.
The treewidth $tw(G)$ of a graph $G$  is the  minimum width among all possible tree decomposition of $G$.
\end{definition} 

\begin{definition}\rm 
    If the tree $T$ of a tree decomposition is a path, then we say that the tree decomposition 
    is a {\it path decomposition}, and similarly the {\it pathwidth}  of a graph $G$  is the  minimum width among all possible path decomposition of $G$.
\end{definition}

A rooted forest is a disjoint union of rooted trees. Given a rooted forest $F$, its \emph{transitive closure} is a graph $H$ in which $V(H)$ contains all the nodes of the rooted forest, and $E(H)$ contain an edge between two vertices only if those two vertices form an ancestor-descendant pair in the forest $F$.

   \begin{definition}
        {\rm  The {\it treedepth} of a graph $G$ is the minimum height of a rooted forest $F$ whose transitive closure contains the graph $G$. It is denoted by $td(G)$.}
    \end{definition}

\subsection{Our Main Results}     
 We show that the {\sc Offensive Alliance} and {\sc Exact Offensive Alliance} problems are W[1]-hard 
parameterized by any of the following parameters: the feedback vertex set number, treewidth, pathwidth, and treedepth
of the input graph.

\subsection{Known Results} The decision version for several types of alliances have been shown to be NP-complete. 
For an integer $r$, a nonempty set $S\subseteq V(G)$ is a {\it defensive $r$-alliance} if for each 
$v\in S$, $|N(v)\cap S|\geq |N(v)\setminus S|+r$. A set is a defensive alliance if it is a defensive 
$(-1)$-alliance. A defensive $r$-alliance $S$ is {\it global} if $S$ is a dominating set. 
 The defensive $r$-alliance problem   is NP-complete for any $r$ \cite{SIGARRETA20091687}. The defensive alliance problem is 
 NP-complete even when restricted to split, chordal and bipartite graph \cite{Lindsay}. 
 For an integer $r$, a nonempty set $S\subseteq V(G)$ is an {\it offensive $r$-alliance} if for each 
$v\in N(S)$, $|N(v)\cap S|\geq |N(v)\setminus S|+r$. An offensive 1-alliance is called an offensive
alliance.  An offensive $r$-alliance $S$ is {\it global} if $S$ is a dominating set. 
 Fernau et al. showed that the offensive $r$-alliance and global 
 offensive $r$-alliance problems are NP-complete for any fixed $r$ \cite{FERNAU2009177}. 
 They also proved that for $r>1$, $r$-offensive alliance is NP-hard, even when restricted to 
 $r$-regular planar graphs.  There are polynomial time algorithms for finding minimum alliances
 in trees \cite{CHANG2012479,Lindsay}.  A polynomial time algorithm for finding minimum defensive alliance in series parallel graph is presented in \cite{10.5555/1292785}. Fernau  and Raible showed in \cite{Fernau} that the defensive and offensive 
 alliance problems and their global
 variants are fixed parameter tractable when parameterized by the solution size $k$. Kiyomi and Otachi
 showed in 
 \cite{KIYOMI201791}, the problems of finding smallest alliances of all kinds are fixed-parameter tractable
 when parameteried by the vertex cover number. The problems of finding smallest defensive 
 and offensive alliances are also fixed-parameter tractable
 when parameteried by the neighbourhood diversity \cite{ICDCIT2021}. 
 Enciso \cite{Enciso2009AlliancesIG} proved that 
 finding defensive and global defensive alliances is fixed parameter tractable when parameterized by domino treewidth. 

\section{Hardness Results}
In this section we show that {\sc Offensive Alliance} is W[1]-hard parameterized by a vertex deletion set to 
trees of height at most seven, that is, a subset $D$ of the vertices of the graph such that 
every component in the graph, after removing $D$, is a tree of height at most seven. 
On the way towards this result, we provide hardness results for several interesting 
versions of the  {\sc Offensive Alliance} problem which  we require in our proofs. 
 The problem 
 {\sc Offensive Alliance$^{\mbox{F}}$} generalizes {\sc Offensive Alliance} 
 where some vertices are 
 forced to be outside the solution; these vertices are called forbidden vertices.  This variant can be formalized as
follows: 
 \vspace{3mm}
    \\
    \fbox
    {\begin{minipage}{33.7em}\label{SP2}
       {\sc Offensive Alliance$^{\mbox{F}}$}\\
        \noindent{\bf Input:} An undirected graph $G=(V,E)$, an integer $r$ and
        a set $V_{\square}\subseteq V(G)$ of forbidden vertices such that each degree one
        forbidden vertex is adjacent to another forbidden vertex and each  forbidden vertex of degree greater than one is adjacent to a degree one forbidden vertex. \\
        \noindent{\bf Question:} Is there an offensive alliance $S\subseteq V$ such that (i) $1\leq |S|\leq r$, and 
    (ii) $S\cap V_{\square}=\emptyset$?
    \end{minipage} }\\

\noindent {\sc  Strong Offensive Alliance$^{\mbox{FN}}$} is a generalization of {\sc  Strong Offensive Alliance$^{\mbox{F}}$} that, in addition, requires
 some ``necessary'' vertices  to be in  $S$.
  This variant can be formalized as
follows: 
\vspace{3mm}
    \\
    \fbox
    {\begin{minipage}{33.7em}\label{SP1}
       {\sc Strong Offensive Alliance$^{\mbox{FN}}$}\\
        \noindent{\bf Input:} An undirected graph $G=(V,E)$, an integer $r$, a set $V_{\triangle}\subseteq V$, and
        a set $V_{\square}\subseteq V(G)$ of forbidden vertices such that  each degree one
        forbidden vertex is adjacent to another forbidden vertex and each  forbidden vertex of degree greater than one is adjacent to a degree one forbidden vertex.  \\
    \noindent{\bf Question:} Is there a strong offensive alliance $S\subseteq V$ such that (i) $1\leq |S|\leq r$, 
    (ii) $S\cap V_{\square}=\emptyset$, and (iii) $V_{\triangle} \subseteq S$?
    \end{minipage} }\\
    
 \noindent While the {\sc  Offensive Alliance}  problem asks for  offensive alliance of 
 size at most $r$,
 we also consider the {\sc Exact  Offensive Alliance} problem that concerns  offensive alliance
 of size exactly $r$. Analogously, we also define exact versions of 
 {\sc Strong Offensive Alliance} presented above. To prove Lemma \ref{oatheorem1}, we consider  the following problem:

\noindent\vspace{3mm}
    \\
    \fbox
    {\begin{minipage}{33.7em}\label{SPM}
       {\sc Multidimensional Relaxed Subset Sum (MRSS)}\\
     \noindent{\bf  Input:} Two integers $k$ and $k'$, a set 
     $S = \{s_1,\ldots,s_n\}$ of vectors with $s_i \in \mathbb{N}^k$ for every $i$ with 
     $1 \leq i \leq  n$  and a target vector $t \in \mathbb{N}^k$.\\
\noindent {\bf Parameter}: $k+k'$ \\
\noindent{\bf Question}: Is there a subset $S'\subseteq S $ with $|S'|\leq k'$ such that $\sum\limits_{s\in S'}{s}\geq t$?
    \end{minipage} }\\
\begin{lemma}\rm \cite{mss}
{\sc MRSS} is W[1]-hard when parameterized by the combined parameter $k+k'$, even if all integers in the input are given in unary.
\end{lemma}
\noindent We now show that the {\sc Strong Offensive Alliance$^{\mbox{FN}}$} problem is W[1]-hard parameterized by  the size of a vertex 
 deletion set into trees of height at most 5, via a reduction from MRSS.

 \begin{lemma}\label{oatheorem1}\rm
 The {\sc Strong Offensive Alliance$^{\mbox{FN}}$} problem is W[1]-hard when parameterized by  the 
 size of a vertex deletion set into trees of height at most 5.
 \end{lemma}
 
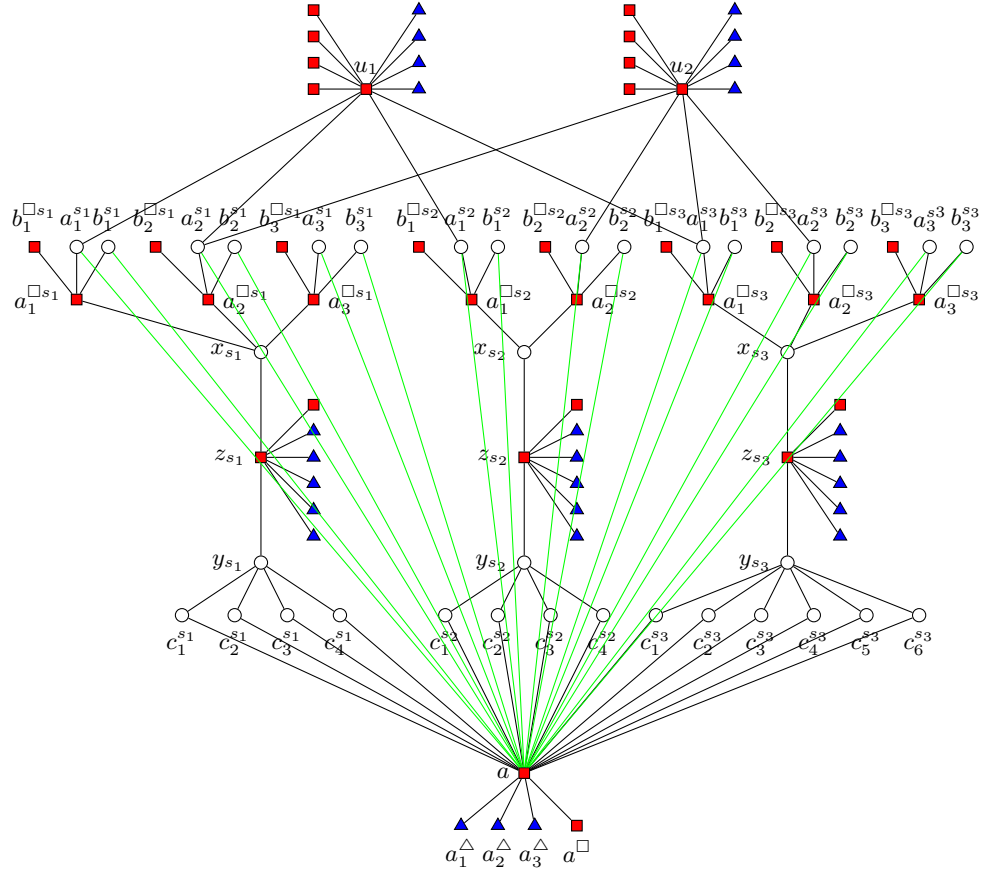
\begin{figure}[ht]
\centering
\begin{tikzpicture}[scale=0.7]
\node[fill=red, square, draw=black, inner sep=0pt, minimum size=0.2cm] (u1) at (-3, 7) [label=above:${u_1}$] {}; 
\node[fill=red, square, draw=black, inner sep=0pt, minimum size=0.2cm] (u11) at (-4, 7) [] {}; 
\node[fill=red, square, draw=black, inner sep=0pt, minimum size=0.2cm] (u12) at (-4, 7.5) [] {}; 
\node[fill=red, square, draw=black, inner sep=0pt, minimum size=0.2cm] (u13) at (-4, 8) [] {}; 
\node[fill=red, square, draw=black, inner sep=0pt, minimum size=0.2cm] (u14) at (-4, 8.5) [] {};

\node[triangle, draw=black, fill= blue, inner sep=0pt, minimum size=0.2cm] (ut11) at (-2,7) [] {};
\node[triangle, draw=black, fill= blue, inner sep=0pt, minimum size=0.2cm] (ut12) at (-2,7.5) [] {};
\node[triangle, draw=black, fill= blue, inner sep=0pt, minimum size=0.2cm] (ut13) at (-2,8) [] {};
\node[triangle, draw=black, fill= blue, inner sep=0pt, minimum size=0.2cm] (ut14) at (-2,8.5) [] {};

\node[fill=red, square, draw=black, inner sep=0pt, minimum size=0.2cm] (u2) at (3, 7) [label=above:${u_2}$] {}; 
\node[fill=red, square, draw=black, inner sep=0pt, minimum size=0.2cm] (u21) at (2, 8) [] {}; 
\node[fill=red, square, draw=black, inner sep=0pt, minimum size=0.2cm] (u22) at (2, 7.5) [] {}; 
\node[fill=red, square, draw=black, inner sep=0pt, minimum size=0.2cm] (u23) at (2, 7) [] {}; 
\node[fill=red, square, draw=black, inner sep=0pt, minimum size=0.2cm] (u24) at (2, 8.5) [] {};

\node[triangle, draw=black, fill= blue, inner sep=0pt, minimum size=0.2cm] (ut1) at (4,7) [] {};
\node[triangle, draw=black, fill= blue, inner sep=0pt, minimum size=0.2cm] (ut2) at (4,7.5) [] {};
\node[triangle, draw=black, fill= blue, inner sep=0pt, minimum size=0.2cm] (ut3) at (4,8) [] {};
\node[triangle, draw=black, fill= blue, inner sep=0pt, minimum size=0.2cm] (ut4) at (4,8.5) [] {};

\node[fill=red, square, draw=black, inner sep=0pt, minimum size=0.2cm] (a1s1) at (-6,3) [label=right:${a_2^{\square s_1}}$] {}; 

\node[fill=red, square, draw=black, inner sep=0pt, minimum size=0.2cm] (a0s1) at (-8.5,3) [label=left:${a_1^{\square s_1}}$] {}; 
\node[fill=red, square, draw=black, inner sep=0pt, minimum size=0.2cm] (b0ss1) at (-9.3,4) [label=above:$b_1^{\square s_1}$] {}; 
\vertex (a0^s1) at (-8.5, 4) [label=above:$a_1^{s_1}$] {}; 
\vertex (b0^s1) at (-7.9, 4) [label=above:$b_1^{s_1}$] {};

\node[fill=red, square, draw=black, inner sep=0pt, minimum size=0.2cm] (b2ss1) at (-4.6,4) [label=above:$b_3^{\square s_1}$] {}; 
\vertex (a2^s1) at (-3.9, 4) [label=above:$a_3^{s_1}$] {}; 
\vertex (b2^s1) at (-3.1, 4) [label=above:$b_3^{s_1}$] {}; 

\node[fill=red, square, draw=black, inner sep=0pt, minimum size=0.2cm] (a2s1) at (-4,3) [label=right:${a_3^{\square s_1}}$] {};   
\node[fill=red, square, draw=black, inner sep=0pt, minimum size=0.2cm] (b1ss1) at (-7,4) [label=above:$b_2^{\square s_1}$] {}; 
\vertex (a1^s1) at (-6.2, 4) [label=above:$a_2^{s_1}$] {}; 
\vertex (b1^s1) at (-5.5, 4) [label=above:$b_2^{s_1}$] {}; 

\vertex (xs1) at (-5, 2) [label=left:$x_{s_1}$] {}; 
\node[fill=red, square, draw=black, inner sep=0pt, minimum size=0.2cm] (zs1) at (-5,0) [label=left:${z_{s_1}}$] {}; 
\vertex (ys1) at (-5, -2) [label=left:$y_{s_1}$] {};
\node[fill=red, square, draw=black, inner sep=0pt, minimum size=0.2cm] (s1z0) at (-4,1) {};  
\node[triangle, draw=black, fill= blue, inner sep=0pt, minimum size=0.2cm] (s1z1) at (-4, .5) [] {}; 
\node[triangle, draw=black, fill= blue, inner sep=0pt, minimum size=0.2cm] (s1z2) at (-4, 0) [] {}; 
\node[triangle, draw=black, fill= blue, inner sep=0pt, minimum size=0.2cm] (s1z3) at (-4, -0.5) [] {}; 
\node[triangle, draw=black, fill= blue, inner sep=0pt, minimum size=0.2cm] (s1z4) at (-4, -1) [] {}; 
\node[triangle, draw=black, fill= blue, inner sep=0pt, minimum size=0.2cm] (s1z5) at (-4, -1.5) [] {}; 

\vertex (c1s1) at (-6.5, -3) [label=below:$c_1^{s_1}$] {};
\vertex (c2s1) at (-5.5, -3) [label=below:$c_2^{s_1}$] {};
\vertex (c3s1) at (-4.5, -3) [label=below:$c_3^{s_1}$] {};
\vertex (c4s1) at (-3.5, -3) [label=below:$c_4^{s_1}$] {};

\node[fill=red, square, draw=black, inner sep=0pt, minimum size=0.2cm] (a) at (0,-6) [label= left:${a}$] {};

\node[triangle, draw=black, fill= blue, inner sep=0pt, minimum size=0.2cm] (a0) at (-1.2,-7) [label=below:${a_1^{\triangle}}$] {};
\node[triangle, draw=black, fill= blue, inner sep=0pt, minimum size=0.2cm] (a1) at (-0.5,-7) [label=below:${a_2^{\triangle}}$] {};
\node[triangle, draw=black, fill= blue, inner sep=0pt, minimum size=0.2cm] (a2) at (0.2,-7) [label=below:${a_3^{\triangle}}$] {};
\node[fill=red, square, draw=black, inner sep=0pt, minimum size=0.2cm] (a3) at (1,-7) [label=below:${a^{\square}}$] {};

\path
(a) edge (a1)
(a) edge (a2)
(a) edge (a3)
(a) edge (a0)
(xs1) edge (a0s1)
(xs1) edge (zs1)
(ys1) edge (zs1)
(xs1) edge (a1s1)
(xs1) edge (a2s1)
(zs1) edge (s1z0)
(zs1) edge (s1z1)
(zs1) edge (s1z2)
(zs1) edge (s1z3)
(zs1) edge (s1z4)
(zs1) edge (s1z5)
(ys1) edge (c1s1)
(ys1) edge (c2s1)
(ys1) edge (c3s1)
(ys1) edge (c4s1)
(a1s1) edge (b1ss1)
(a1s1) edge (b1^s1)
(a1s1) edge (a1^s1)
(a2s1) edge (b2ss1)
(a2s1) edge (b2^s1)
(a2s1) edge (a2^s1)
(u1) edge (u11)
(u1) edge (u12)
(u1) edge (u13)
(u1) edge (u14)
(u2) edge (u21)
(u2) edge (u22)
(u2) edge (u23)
(u2) edge (u24)
(u2) edge (ut1)
(u2) edge (ut2)
(u2) edge (ut3)
(u2) edge (ut4)
(u1) edge (ut11)
(u1) edge (ut12)
(u1) edge (ut13)
(u1) edge (ut14)
(a0s1) edge (b0ss1)
(a0s1) edge (a0^s1)
(a0s1) edge (b0^s1);

\node[fill=red, square, draw=black, inner sep=0pt, minimum size=0.2cm] (a1s2) at (-1,3) [label=right:${a_1^{\square s_2}}$] {};

\node[fill=red, square, draw=black, inner sep=0pt, minimum size=0.2cm] (a2s2) at (1,3) [label=right:${a_2^{\square s_2}}$] {}; 

\node[fill=red, square, draw=black, inner sep=0pt, minimum size=0.2cm] (b1ss2) at (-2,4) [label=above:$b_1^{\square s_2}$] {}; 
\vertex (a1^s2) at (-1.2, 4) [label=above:$a_1^{s_2}$] {}; 
\vertex (b1^s2) at (-0.5, 4) [label=above:$b_1^{s_2}$] {}; 

\node[fill=red, square, draw=black, inner sep=0pt, minimum size=0.2cm] (b2ss2) at (0.4,4) [label=above:$b_2^{\square s_2}$] {}; 
\vertex (a2^s2) at (1.1, 4) [label=above:$a_2^{s_2}$] {}; 
\vertex (b2^s2) at (1.9, 4) [label=above:$b_2^{s_2}$] {};

\node[fill=red, square, draw=black, inner sep=0pt, minimum size=0.2cm] (zs2) at (0,0) [label=left:${z_{s_2}}$] {}; 
\vertex (xs2) at (0, 2) [label=left:$x_{s_2}$] {};
\vertex (ys2) at (0, -2) [label=left:$y_{s_2}$] {};

\node[fill=red, square, draw=black, inner sep=0pt, minimum size=0.2cm] (s2z0) at (1,1) {};  
\node[triangle, draw=black, fill= blue, inner sep=0pt, minimum size=0.2cm] (s2z1) at (1, .5) [] {}; 
\node[triangle, draw=black, fill= blue, inner sep=0pt, minimum size=0.2cm] (s2z2) at (1, 0) [] {}; 
\node[triangle, draw=black, fill= blue, inner sep=0pt, minimum size=0.2cm] (s2z3) at (1, -0.5) [] {}; 
\node[triangle, draw=black, fill= blue, inner sep=0pt, minimum size=0.2cm] (s2z4) at (1, -1) [] {}; 
\node[triangle, draw=black, fill= blue, inner sep=0pt, minimum size=0.2cm] (s2z5) at (1, -1.5) [] {}; 
\vertex (c1s2) at (-1.5, -3) [label=below:$c_1^{s_2}$] {};
\vertex (c2s2) at (-0.5, -3) [label=below:$c_2^{s_2}$] {};
\vertex (c3s2) at (0.5, -3) [label=below:$c_3^{s_2}$] {};
\vertex (c4s2) at (1.5, -3) [label=below:$c_4^{s_2}$] {};

\path
(xs2) edge (zs2)
(ys2) edge (zs2)
(xs2) edge (a1s2)
(xs2) edge (a2s2)
(zs2) edge (s2z0)
(zs2) edge (s2z1)
(zs2) edge (s2z2)
(zs2) edge (s2z3)
(zs2) edge (s2z4)
(zs2) edge (s2z5)
(ys2) edge (c1s2)
(ys2) edge (c2s2)
(ys2) edge (c3s2)
(ys2) edge (c4s2)
(a1s2) edge (b1ss2)
(a1s2) edge (b1^s2)
(a1s2) edge (a1^s2)
(a2s2) edge (b2ss2)
(a2s2) edge (b2^s2)
(a2s2) edge (a2^s2);

\node[fill=red, square, draw=black, inner sep=0pt, minimum size=0.2cm] (a1s3) at (3.5,3) [label=right:${a_1^{\square s_3}}$] {}; 
\node[fill=red, square, draw=black, inner sep=0pt, minimum size=0.2cm] (a2s3) at (5.5,3) [label=right:${a_2^{\square s_3}}$] {}; 
\node[fill=red, square, draw=black, inner sep=0pt, minimum size=0.2cm] (a3s3) at (7.5,3) [label=right:${a_3^{\square s_3}}$] {}; 

\node[fill=red, square, draw=black, inner sep=0pt, minimum size=0.2cm] (b1ss3) at (2.7,4) [label=above:$b_1^{\square s_3}$] {}; 
\vertex (a1^s3) at (3.4, 4) [label=above:$a_1^{s_3}$] {}; 
\vertex (b1^s3) at (4, 4) [label=above:$b_1^{s_3}$] {}; 

\node[fill=red, square, draw=black, inner sep=0pt, minimum size=0.2cm] (b2ss3) at (4.8,4) [label=above:$b_2^{\square s_3}$] {}; 
\vertex (a2^s3) at (5.5, 4) [label=above:$a_2^{s_3}$] {}; 
\vertex (b2^s3) at (6.2, 4) [label=above:$b_2^{s_3}$] {}; 

\node[fill=red, square, draw=black, inner sep=0pt, minimum size=0.2cm] (b3ss3) at (7,4) [label=above:$b_3^{\square s_3}$] {}; 
\vertex (a3^s3) at (7.7, 4) [label=above:$a_3^{s3}$] {}; 
\vertex (b3^s3) at (8.4, 4) [label=above:$b_3^{s_3}$] {}; 

\node[fill=red, square, draw=black, inner sep=0pt, minimum size=0.2cm] (zs3) at (5,0) [label=left:${z_{s_3}}$] {}; 
\vertex (xs3) at (5, 2) [label=left:$x_{s_3}$] {};
\vertex (ys3) at (5, -2) [label=left:$y_{s_3}$] {};
\node[fill=red, square, draw=black, inner sep=0pt, minimum size=0.2cm] (s3z0) at (6,1) {};  
\node[triangle, draw=black, fill= blue, inner sep=0pt, minimum size=0.2cm] (s3z1) at (6, .5) [] {}; 
\node[triangle, draw=black, fill= blue, inner sep=0pt, minimum size=0.2cm] (s3z2) at (6, 0) [] {}; 
\node[triangle, draw=black, fill= blue, inner sep=0pt, minimum size=0.2cm](s3z3) at (6, -0.5) [] {}; 
\node[triangle, draw=black, fill= blue, inner sep=0pt, minimum size=0.2cm] (s3z4) at (6, -1) [] {}; 
\node[triangle, draw=black, fill= blue, inner sep=0pt, minimum size=0.2cm] (s3z5) at (6, -1.5) [] {}; 
\vertex (c1s3) at (2.5, -3) [label=below:$c_1^{s_3}$] {};
\vertex (c2s3) at (3.5, -3) [label=below:$c_2^{s_3}$] {};
\vertex (c3s3) at (4.5, -3) [label=below:$c_3^{s_3}$] {};
\vertex (c4s3) at (5.5, -3) [label=below:$c_4^{s_3}$] {};
\vertex (c5s3) at (6.5, -3) [label=below:$c_5^{s_3}$] {};
\vertex (c6s3) at (7.5, -3) [label=below:$c_6^{s_3}$] {};
\path
(xs3) edge (zs3)
(ys3) edge (zs3)
(xs3) edge (a1s3)
(xs3) edge (a2s3)
(xs3) edge (a3s3)
(zs3) edge (s3z0)
(zs3) edge (s3z1)
(zs3) edge (s3z2)
(zs3) edge (s3z3)
(zs3) edge (s3z4)
(zs3) edge (s3z5)
(ys3) edge (c1s3)
(ys3) edge (c2s3)
(ys3) edge (c3s3)
(ys3) edge (c4s3)
(ys3) edge (c5s3)
(ys3) edge (c6s3)
(a1s3) edge (b1ss3)
(a1s3) edge (b1^s3)
(a1s3) edge (a1^s3)
(a2s3) edge (b2ss3)
(a2s3) edge (b2^s3)
(a2s3) edge (a2^s3)
(a3s3) edge (b3ss3)
(a3s3) edge (b3^s3)
(a3s3) edge (a3^s3)
(u1) edge (a1^s1)
(u1) edge (a0^s1)
(u1) edge (a1^s2)
(u1) edge (a1^s3)
(u2) edge (a1^s1)
(u2) edge (a2^s2)
(u2) edge (a1^s3)
(u2) edge (a2^s3)
(a) edge (c1s1)
(a) edge (c2s1)
(a) edge (c3s1)
(a) edge (c4s1)
(a) edge (c1s2)
(a) edge (c2s2)
(a) edge (c3s2)
(a) edge (c4s2)
(a) edge (c1s3)
(a) edge (c2s3)
(a) edge (c3s3)
(a) edge (c4s3)
(a) edge (c5s3)
(a) edge (c6s3);

\path 
(a0^s1) [green] edge (a)
(b0^s1) [green] edge (a)
(a1^s1) [green] edge (a)
(b1^s1) [green] edge (a)
(a2^s1) [green] edge (a)
(b2^s1) [green] edge (a)
(a2^s2) [green] edge (a)
(b2^s2) [green] edge (a)
(a1^s2) [green] edge (a)
(b1^s2) [green] edge (a)
(a1^s3) [green] edge (a)
(b1^s3) [green] edge (a)
(a2^s3) [green] edge (a)
(b2^s3) [green] edge (a)
(a3^s3) [green] edge (a)
(b3^s3) [green] edge (a);

 \end{tikzpicture}
 \caption{The graph $G$ in the proof of Lemma \ref{oatheorem1} constructed for MRSS instance 
 $S=\{(2, 1), (1,1), (1,2)\}$, $t=(3,3)$, $k=2$ and $k'=2$.}
     \label{oafig1}
\end{figure}

 \proof To prove this we reduce from MRSS, which  is known to be W[1]-hard when 
parameterized by the combined parameter $k+k'$, even if all integers in the input are given in unary \cite{mss}.
Let  $I = (k, k', S, t)$  be an instance of {\sc MRSS}. We construct an  instance 
 $I'=(G,r,V_{\triangle},V_{\square})$ of {\sc Strong Offensive Alliance$^{\mbox{FN}}$} in the following way.
 See Figure \ref{oafig1} for an illustration.
 \noindent First, we  introduce a  set  of $k$ forbidden vertices $U=\{ u_1,u_2,\ldots,u_k\}$. For each vector
$s=(s(1),s(2),\ldots,s(k))\in S$, we introduce a tree $T_{s}$ into $G$.  We define $\mbox{max}(s)=\max\limits_{1\leq i \leq k}\{s(i)\}$.
The vertex set of tree $T_s$ is defined as follows:
$$ V(T_s) = A_s \cup B_s \cup A^{\square}_s \cup B^{\square}_s \cup C_{s} \cup Z_{s}  
\cup \Big\{x_s,y_s,z_s\Big\} $$
where  $A_s=\{a^{s}_{1},\ldots,a^s_{\max(s)+1}\}$, 
$B_{s}=\{b^{ s}_{1},\ldots,b^{ s}_{\max(s)+1}\}$, 
$ A_{s}^{\square}=\{a^{\square s}_{1},\ldots,a^{\square s }_{\max(s)+1}\}$, 
$B_s^{\square}=\{b^{ \square s}_{1},\ldots,b^{\square s }_{\max(s)+1}\}$ and 
$C_{s}=\{c_{1}^{s},\ldots,c_{2\max{(s)+2}}^{s}\}$ are five sets of vertices, and 
the set $Z_{s}=\{z^{\triangle s}_{1},z^{\triangle s}_{2},z^{\triangle s}_{3},
z^{\triangle s}_{4},z^{\triangle s}_{5},
z^{\square s} \}$ contains five necessary vertices and one forbidden vertex. 
We now create the edge set of $T_s$.
\begin{align*}
    E(T_s) &=  \bigcup\limits_{i=1}^{{\max(s)+1}} 
    \Big\{ (a_{i}^{\square s},b_{i}^{\square s}), (a_{i}^{\square s},a_{i}^{s}),
    (a_{i}^{\square s},b_{i}^{s}) ,(x_s,a_{i}^{\square s})\Big\} \\ 
    & \bigcup\limits_{i=1}^{5} \{(z_s,z_i^{\triangle s}),(z_s,z^{\square s})\} 
    \bigcup \{(x_s,z_s),(z_s,y_s)\} 
     \bigcup\limits_{i=1}^{2\max(s)+2} (y_s,c_i^s)
\end{align*}

\noindent  Next we introduce a vertex $a$ and a set of four vertices $A=\{a_{1}^{\triangle},a_{2}^{\triangle},a_{3}^{\triangle},
 a^{\square}\}$ containing three necessary vertices and one forbidden vertex. Make $a$ adjacent to 
 all the vertices in $A$.
  For each $i\in \{1,2,\ldots,k\}$ and 
 for each $s\in S$, we make $u_{i}$ adjacent to exactly $s(i)$ many vertices 
 of  $A_s$ in arbitrary manner.  For each $s\in S$, we make $a$ adjacent to all  
 the vertices of  $A_s\cup B_{s} \cup  C_s$. 
 For every $u_i\in U$, we create a set $V_{u_i\square}$ of $\sum\limits_{s\in S}{s(i)}$ 
 forbidden vertices  and a set $V_{u_i\triangle}$ of $2\sum\limits_{s\in S}{s(i)}-2t(i)+2$ 
 necessary vertices; and make $u_i$ adjacent to every vertex 
 of $V_{u_i\square}\cup V_{u_i\triangle}$.
We define $$V_{\triangle} = \bigcup\limits_{i=1}^{k} V_{u_i\triangle} \bigcup A \setminus \{a^{\square}\} 
\bigcup\limits_{s\in S} {Z_{s} \setminus \{ z^{\square s}\} }   $$ and 
$$V_{\square} = U\cup \{a,a^{\square}\} \bigcup\limits_{i=1}^{k} V_{u_i\square} \bigcup\limits_{s\in S} 
A_s^{\square}\cup B_s^{\square}\cup \{z_s, z^{\square s}\}.$$
We set $r= \sum\limits_{i=1}^{k} {2\Big(\sum\limits_{s\in S}{s(i)}-t(i)+1\Big)}+\sum\limits_{s\in S}2(\max(s)+1)+5n+3+k'$.
Observe that if we remove the set $U\cup \{a\}$ of $k+1$ vertices from $G$, each 
connected component 
of the resulting graph is a tree with height at most 5. Note that, $I'$ can be constructed in polynomial time. The reason is this. As 
all integers in $I$ are bounded by a polynomial in $n$, the number of vertices in $G$ is also polynomially  bounded in $n$.

\par It remains to show that  $I$ is a yes instance if and only if $I'$ is a yes instance. 
Towards showing the forward direction, let $S'$ be a subset of $S$ such that $|S'|\leq k'$ and $\sum\limits_{s\in S'}{s}\geq t$. 
We claim 
$$ R = V_{\triangle} \bigcup\limits_{s\in S'} A_{s}\cup B_{s}\cup \{x_{s}\} \bigcup\limits_{s\in S\setminus S'} C_{s} $$
is a strong offensive alliance of $G$ such that $|R|\leq r$, $V_{\triangle}\subseteq R$, and $V_{\square} \cap R=\emptyset$. 
Observe that 
$N_{G}(R) = U \cup \{a\} \bigcup\limits_{s\in S} \{z_s\} \bigcup\limits_{s\in S\setminus S'} \{y_{s}\} \bigcup\limits_{s\in S'} A_{s}^{\square} $. 
Let $u_i \in U$, 
then we show that $d_{R}(u_{i})\geq d_{R^c}(u_{i})+2$. 
As $\sum\limits_{s \in S'} s(i)-t(i)\geq 0$, we get
\begin{equation*}
\begin{split}
 d_{R}(u_i)&=\sum\limits_{s \in S'} s(i) + |V_{u_i\triangle}|\\
           &=\sum\limits_{s \in S'} s(i)+2\sum\limits_{s\in S}{s(i)}- 2t(i)+2\\
           &=\Big(\sum\limits_{s \in S'} s(i)-t(i)\Big )+ \sum\limits_{s\in S}{s(i)}-t(i)+ \sum\limits_{s\in S}{s(i)}+2 \\
           &\geq \sum\limits_{s\in S}{s(i)} -t(i) + \sum\limits_{s\in S}{s(i)}+2 \\
           &=\sum\limits_{s\in S \setminus S'}{s(i)} +\Big(\sum\limits_{s\in S'}{s(i)} -t(i) \Big)+ \sum\limits_{s\in S}{s(i)} +2\\
           &\geq  \sum\limits_{s\in S \setminus S'}{s(i)}+\sum\limits_{s\in S}{s(i)}+2
           =\sum\limits_{s\in S \setminus S'}{s(i)}+|V_{u_i\square}|+2\\
           &=d_{R^c}(u_i)+2.
 \end{split}          
\end{equation*}
For the remaining vertices $x$ in $N(R)$, 
it is easy to see that $d_{R}(x) \geq d_{R^c}(x)+2$. 
Therefore, $R$ is a strong offensive alliance.

\par Towards showing the reverse direction of the equivalence, suppose 
$G$ has a strong offensive alliance 
$R$ of size at most $r$  such that $V_{\triangle} \subseteq R$ 
and $V_{\square}\cap R=\emptyset$. From the definition of $V_{\triangle}$
and $V_{\square}$, it is easy to note that $U \subseteq N(R)$. 
We know $V_{\triangle}$ contains $\sum\limits_{i=1}^{k} {\Big(\sum\limits_{s\in S}{2s(i)}-2t(i)\Big)+5n+3}$ vertices; thus
besides the vertices of $V_{\triangle}$, there are 
at most $\sum\limits_{s\in S}2(\max(s)+1)+k'$  vertices in $R$.
Since $a\in N(V_{\triangle})$ and $d_{G}(a) = \sum\limits_{s\in S}4(\max(s)+1) +4$ where $a$ is adjacent to three necessary vertices, it must have at least $\sum\limits_{s\in S}2(\max(s)+1)$ many neighbours 
in $R$ from the set $\bigcup\limits_{s\in S} (A_{s}\cup B_{s}\cup C_s)$. 
It is to be noted that if a vertex from the set $A_{s}\cup B_s$ 
is in the solution  then the whole set $A_{s}\cup B_s\cup \{x_s\}$ lie in the solution. 
Otherwise  $v\in A_{s}^{\square}\subseteq N(R)$ will have $d_R(v)<d_{R^c}(v)+2$ 
which is a contradiction as $R$ is a strong offensive alliance. 
This shows that at most $k'$ many sets of the form $A_s\cup B_s\cup \{x_s\}$  contribute to the solution as 
otherwise the size of solution  exceeds $r$. 
Therefore, any  strong offensive alliance $R$ of size at most 
$r$  can be transformed to another strong offensive alliance $R'$ of size at most $r$ as follows:
$$ R' =  V_{\triangle} \bigcup\limits_{x_{s}\in R} A_{s}\cup B_s\cup \{x_{s}\} \bigcup\limits_{x_{s}\in V(G)\setminus R} C_{s}.$$ 
We define a subset $S'=\Big\{ s \in S ~|~ x_s\in R'\Big\}.$
Clearly, $|S'|\leq k'$. We claim that $\sum\limits_{s\in S'} s(i) \geq t(i)$ for all $1 \leq i \leq k$. 
 Assume for the sake of contradiction that $\sum\limits_{s\in S'} s(i) < t(i)$ for some $i \in \{1,2,\ldots,k\}$. 
Then, we have 
\begin{equation*}
\begin{split}
 d_{R'}(u_i)&=\sum\limits_{s \in S'} {s(i)} +|V_{u_i\triangle}|\\
           &=\sum\limits_{s \in S'} {s(i)}+2\sum\limits_{s\in S}{s(i)}- 2t(i)+2\\
           &=\sum\limits_{s \in S'} {s(i)}-t(i) +\sum\limits_{s\in S}{s(i)}- t(i) +\sum\limits_{s\in S}{s(i)}+2\\
           &< \sum\limits_{s\in S}{s(i)} -t(i) +\sum\limits_{s\in S}{s(i)} +2\\
           &= \sum\limits_{s\in S\setminus S'}{s(i)}+\Big(\sum\limits_{s\in S'}{s(i)} -t(i)\Big) +\sum\limits_{s\in S}{s(i)}+2\\
           &<  \sum\limits_{s\in S \setminus S'}{s(i)} +\sum\limits_{s\in S}{s(i)}+2=\sum\limits_{s\in S \setminus S'}{s(i)} +|V_{u_i\square}|+2\\
           &=d_{R'^c}(u_i)+2
 \end{split}          
\end{equation*}
and we also know  $u_{i}\in N(R')$, which is a contradiction to the fact that $R'$ is a strong offensive alliance. This shows that $I$ is a yes instance. \\

\begin{figure}[H]
\centering
\[\begin{tikzpicture}[scale=0.5][H]

\centering

\node [fill=red, square, draw=black, inner sep=0pt, minimum size=0.15cm] (a) at (0, .5) [label=left:$x$]{};

\node [triangle, draw=black, fill= yellow, inner sep=0pt, minimum size=0.2cm] (y) at (1.5, 0.5) [label=right:$y$]{};

\node[fill=red, square, draw=black, inner sep=0pt, minimum size=0.15cm](a1) at (-2, -0.5) []{};
\node[fill=red, square, draw=black, inner sep=0pt, minimum size=0.15cm](b7) at (-1, -0.5) []{};
\node[fill=red, square, draw=black, inner sep=0pt, minimum size=0.15cm] (a2) at (2, -0.5) []{};
\node [triangle, draw=black, fill= blue, inner sep=0pt, minimum size=0.2cm] (b1) at (-2, 2) [label=above:$v_{1}$]{};
\node [triangle, draw=black, fill= blue, inner sep=0pt, minimum size=0.2cm] (b5) at (-1, 2) [label=above:$v_{2}$]{};
\node [triangle, draw=black, fill= blue, inner sep=0pt, minimum size=0.2cm] (b2) at (2, 2) [label=above:$v_{\ell}$]{};

\node (t) at (0, -1) [label=below:$V_{x}^{\square}$]{};

\path 
(a) edge (y)
(a) edge (b7)
(a) edge (b5)
(a) edge (a1)
(a) edge (a2)
(a) edge (b1)
(a) edge (b2);

\draw[dotted, thick] (-0.2,-0.5) -- (1.3,-0.5);
\draw[dotted, thick] (-.2,2) -- (1.3,2);

\draw [decorate,decoration={brace,amplitude=10pt,mirror},xshift=0pt,yshift=0pt] (-2,-0.7) --  (2,-0.7) node [black,midway,xshift=-0.6cm]{};

\end{tikzpicture}
\]
\caption{An illustration of the gadget used in the proof of Corollary \ref{corollary}.}
\label{CorollaryFig1}
\end{figure}
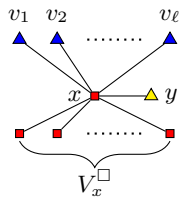

\noindent We have the following corollaries  from Lemma \ref{oatheorem1}.
 \begin{corollary}\label{corollary}\rm
 The {\sc Strong Offensive Alliance$^{\mbox{FN}}$} problem is W[1]-hard when parameterized by  the size of 
 a vertex deletion set into trees of height at most 5, even when $|V_{\triangle}|=1$. 
 \end{corollary}
  \proof Given an instance $I=(G,r,V_{\triangle},V_{\square})$ of {\sc Strong  Offensive Alliance$^{\mbox{FN}}$},
 we construct an equivalent instance $I'=(G',r',V_{\triangle}^{\prime},V_{\square}^{\prime})$ 
 with $|V_{\triangle}^{\prime}|=1$. See Figure \ref{CorollaryFig1} for an illustration. 
\noindent Let $v_1,v_2,\dots,v_{\ell}$ be vertices of $V_{\triangle}$ where we assume that $\ell>1$.
 We introduce two vertices $x$ and $y$ where $x$ is a forbidden vertex and $y$ is a necessary vertex;
 and make $x$ and $y$ adjacent. 
 We make $x$ adjacent to all the vertices in $V_{\triangle}$. 
 We also introduce a set $V_{x\square}$ of $\ell-1$  forbidden vertices and make them  adjacent to $x$. 
Set $r'=r+1$.  Define $V_{\triangle}^{\prime}=\{y\}$ and $V_{\square}^{\prime}=\{x \}\cup V_{x\square} \cup V_{\square}$. 
 We also define $G'$ as follows $$ V(G') = V(G) \cup \{x,y\} \cup V_{x\square} $$ and $$ E(G') = E(G) \bigcup \Big\{(x,y),(x,\alpha),(x,\beta) ~|~ \alpha\in V_{x\square},\beta \in V_{\triangle} \Big\}.$$  
 Let $H$  be a vertex deletion set   of $G$ into trees of height at most 5. 
 Clearly, if $H$ has at most $k$ vertices then the set $H\cup \{x\}$ has at most $k+1$ vertices and  is a vertex deletion set
 of $G'$ into trees of height at most 5. 
 It is easy to see that $I$ and $I'$ are equivalent  instances.\qed\\\\
 
 \noindent We can get an analogous result for the exact variant.
 \begin{corollary}\label{corollaryExactSOAFN}\rm
 The {\sc Exact Strong Offensive Alliance$^{\mbox{FN}}$} problem is W[1]-hard when parameterized by  the size of a vertex deletion set into trees of height at most 5 even when $|V_{\triangle}|=1$.
 \end{corollary}
 
\noindent  Next, we give an FPT reduction that eliminates necessary vertices.  
 
\begin{figure}[ht]
\centering
 \begin{tikzpicture}[scale=0.3]
 
\draw[green, dashed, thick] (-7,7) -- (9,7);
\draw[green, dashed, thick] (-7,7) -- (-7,4);
\draw[green, dashed, thick] (-7,4) -- (9,4);
\draw[green, dashed, thick] (9,7) -- (9,4);
 
\node[fill=red, square, draw=black, inner sep=0pt, minimum size=0.15cm] (x00) at (0,0) [label= right:${x^{\square}}$] {};

\vertex (v1) at (-2,5) [label=above:${v_{1}}$] {};
\vertex (v2) at (0,5) [label=above:${v_{2}}$] {};
\vertex (v3) at (3,5) [label=above:${v_{n'}}$] {};

\node[fill=red, square, draw=black, inner sep=0pt, minimum size=0.15cm] (f1) at (5,5) [label=above:${f_{1}}$] {};
\node[fill=red, square, draw=black, inner sep=0pt, minimum size=0.15cm] (f2) at (8,5) [label=above:${f_{n''}}$] {};

\draw[dotted, thick] (5.5,5) -- (7.5,5);
\draw[dotted, thick] (0.5,5) -- (2.5,5);

\draw[dotted, thick] (-5,-3) -- (-5,-1);

\node (g) at (9,5) [label=right:${G}$] {};

\node[fill=red, square, draw=black, inner sep=0pt, minimum size=0.15cm] (x01) at (0,-2) [] {};

\draw[dotted, thick] (0.3,-2) -- (1.7,-2);

\node[fill=red, square, draw=black, inner sep=0pt, minimum size=0.15cm] (x02) at (2,-2) [] {};

\node (vx) at (1,-2) [label= below:$V_{x}^{\square}$] {};

\node [triangle, draw=black, fill= blue, inner sep=0pt, minimum size=0.15cm] (x) at (-5,5) [label= above:${x}$] {};

\vertex (y1) at (-5,2) [label= left:${t_{1}}$] {};
\vertex (y2) at (-5,0) [label= left:${t_{2}}$] {};
\vertex (yn) at (-5,-4) [label= left:${t_{4n}}$] {};

\node[fill=red, square, draw=black, inner sep=0pt, minimum size=0.15cm] (t) at (-7.5,1) [label=above:${t^{\square}}$] {};

\node[fill=red, square, draw=black, inner sep=0pt, minimum size=0.15cm] (t0) at (-9.5,2) [] {};
\node[fill=red, square, draw=black, inner sep=0pt, minimum size=0.15cm] (t1) at (-9.5,0) [] {};

\node (T) at (-9.5,1) [label=left:${V_{t}^{\square}}$] {};

\draw[dotted, thick] (-9.5,1.8) -- (-9.5,0.2);

\path 

(x00) edge (x01)
(x00) edge (x02)
(x00) edge (x)
(x00) edge (v1)
(x00) edge (v2)
(x00) edge (v3)
(x00) edge (y1)
(x00) edge (y2)
(x00) edge (yn)
(t) edge (x)
(t) edge (y1)
(t) edge (y2)
(t) edge (yn)
(t) edge (t0)
(t) edge (t1);

\end{tikzpicture}
    \caption{The reduction from {\sc Strong Offensive Alliance$^{\mbox{FN}}$} to 
    {\sc Offensive Alliance$^{\mbox{F}}$} in Lemma \ref{twtheorem3}. Note that the set $\{v_{1},\ldots,v_{n'}\}$ may contain forbidden vertices of degree greater  than one.}
    \label{oafig2}
\end{figure}
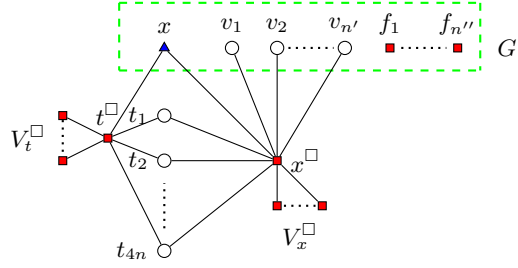

\begin{lemma}\label{twtheorem3}\rm
 The {\sc Offensive Alliance$^{\mbox{F}}$} problem is W[1]-hard when parameterized by the size of a vertex deletion set into trees of height at most 5.
 \end{lemma}
 \proof  To prove this we reduce from the {\sc Strong Offensive Alliance$^{\mbox{FN}}$} problem, which is 
 W[1]-hard when parameterized by  the size of a vertex deletion set into trees of height at most 5, even when $|V_{\triangle}|=1$.
 See Corollary \ref{corollary}. 
Given an instance $I=(G,r,V_{\triangle}=\{x\},V_{\square})$ of {\sc Strong Offensive Alliance$^{\mbox{FN}}$}, 
 we construct an instance $I'=(G',r',V^{\prime}_{\square})$ of {\sc Offensive Alliance$^{\mbox{F}}$} 
 the following way. See Figure \ref{oafig2} for an illustration.
 Let $n$ be the number of vertices in $G$ and 
 let $V(G)=\{x, v_1, v_2,\ldots, v_{n'},f_{1},\ldots,f_{n''}\}$ 
 where $F=\{f_{1},\ldots,f_{n''}\}$ is the set of  degree one forbidden vertices in $V(G)$. 
 We introduce two vertices $t^{\square},x^{\square}$ into $G'$. We create a set 
 $V_{t}^{\square}=\{t^{\square}_{1},\ldots,t^{\square}_{4n}\}$
 of $4n$  forbidden vertices into $G'$ and make them adjacent to $t^{\square}$. We  
 introduce a set 
$V_{x}^{\square}$  of $n$  forbidden vertices and make them  adjacent to
  $x^{\square}$. 
Finally we create a set $T=\{t_{1},\ldots,t_{4n}\}$ of $4n$ vertices and 
make the vertices in $T$ adjacent to $t^{\square}$ and $x^{\square}$,
and make the vertices in $V(G)\setminus F$ adjacent to $x^{\square}$. 
We also add an edge $(x,t^{\square})$.
 Set $r'=r+4n$.  We define $G'$ as follows:
 $$ V(G') = V(G) \cup T \cup V_{t}^{\square} \cup V_{x}^{\square} \cup \{t^{\square},x^{\square}\}  $$ and 
\begin{align*}
    E(G') =& E(G)  
    \cup  \Big \{(t^{\square},\alpha)~:~\alpha \in T \cup V_t^{\square} \cup\{x\}\Big\} \\
    & \cup  \Big\{(x^{\square},\beta)~:~\beta \in T\cup V_{x}^{\square} \cup V(G)\setminus F \Big\}
\end{align*}

\noindent  We define $V_{\square}^{\prime} = V_{\square} \cup  V_{t}^{\square} \cup V_{x}^{\square} \cup \{t^{\square},x^{\square}\}$. Observe that there exists a set of at most $k+2$ vertices in $G'$  
whose deletion makes the resulting graph a forest containing trees of height at most 5.  
We can find  such a 
set because there exists a vertex deletion set $H$ of $G$ into trees of height at most 5.  
We  just add $\{x^{\square},t^{\square}\}$ to the set $H$, then the resulting set is of size $k+2$ whose deletion makes the resulting graph a forest containing trees of height at most 5.

 \par We  now claim  that $I$ is a yes-instance if and only if $I'$ is a yes-instance. 
Assume first  that $R$ is a strong offensive alliance of size at most
$r$ in $G$ such that $\{x\} \subseteq R$ and $V_{\square}\cap R=\emptyset$. 
We claim  $R'=R \cup T$ is an offensive alliance of size at most $r+4n$ in $G'$ such that $V^{\prime}_{\square} \cap R'=\emptyset$. Clearly, $N(R') = \{t^{\square},x^{\square}\} \cup N(R)$. For each $v\in N(R)$, 
we know that $d_{R}(v)\geq d_{R^c}(v)+2$ in $G$. Therefore in graph $G'$, 
we get $d_{R'}(v)\geq d_{R'^c}(v)+1$ for each $v\in N(R)$ due to the vertex $x^{\square}$. 
For   $v\in \{x^{\square},t^{\square}\}$, it is clear that $d_{R'}(v)\geq d_{ R'^c}(v)+1$.
This shows that $I'$ is a yes instance. 

\par To prove the reverse direction of the equivalence, suppose  
$R'$ is an offensive alliance of size at most $r'=r+4n$ in $G'$ such that 
$R'\cap V_{\square}^{\prime}=\emptyset$. We claim that $x \in R'$. 
In fact, we show that $T\cup \{x\} \subseteq R'$. Since $R'$ is non empty, it must contain 
a vertex from the set $V(G)\cup T$. Then $x^{\square}\in N(R')$. 
Due to $n$ forbidden vertices in the set $V_{x}^{\square}$, node
$x^{\square}$ must have at least $n+1$ neighbours in $R'$. 
This implies that $R'$ contains at least one vertex from $T$.  
Then $t^{\square}\in N(R')$ and it satisfies the condition 
$d_{R'}(t^{\square})\geq d_{{R'}^c}(t^{\square})+1$. Since $|V_{t}^{\square}|=4n$, the
condition $d_{R'}(t^{\square})\geq d_{R'^c}(t^{\square})+1$ forces the set 
$\{x\}\cup T$ to be 
inside the solution. Consider $R=R' \cap V(G)$. Clearly $|R|\leq r$, $x\in R$, 
$R\cap V_{\square}=\emptyset $ and we 
show that $R$ is a strong offensive alliance in $G$. 
For each $v\in N(R') \cap V(G) =N(R)$, we have $N_{R'}(v)\geq N_{R'^c}(v)+1$ in
$G'$.
Notice that we do not have  $x^{\square}$ in $G$ which is  adjacent to all vertices in $N(R)$.  
Thus for each $v\in N(R)$, we get $N_{R}(v)\geq N_{R^c}(v)+2$  in $G$. Therefore $R$ is a strong 
offensive alliance of size at most $r$ in $G$ such that $x\in R$ and 
$R\cap V_{\square}=\emptyset$. This shows that $I$ is a yes instance. \qed \\\\

\noindent We are now ready to show our main hardness result for {\sc Offensive Alliance} using a reduction
from {\sc Offensive Alliance$^{\mbox{F}}$}.

\begin{theorem}\label{twtheorem}\rm
 The {\sc Offensive Alliance} problem is W[1]-hard when parameterized by the size of a vertex deletion set into trees of height at most 7.
 \end{theorem}
 
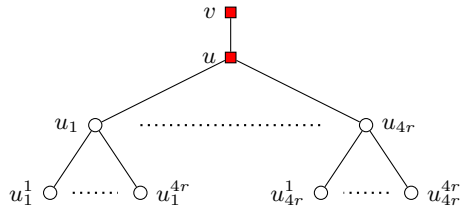
\begin{figure}[ht]
    \centering
 \begin{tikzpicture}[scale=0.6]

\node[fill=red, square, draw=black, inner sep=0pt, minimum size=0.2cm] (h) at (0,0) [label=left:$v$] {};
\node[fill=red, square, draw=black, inner sep=0pt, minimum size=0.2cm] (h1) at (0,-1) [label=left:$u$] {};
\vertex (h11) at (-3,-2.5) [label=left:$u_{1}$] {};
\vertex (h12) at (3,-2.5) [label=right:$u_{4r}$] {};

\vertex (h21) at (-4,-4) [label=left:${u_{1}^{ 1 }}$] {};
\vertex (h22) at (-2,-4) [label=right:${u_{1}^{ 4r }}$] {};

\vertex (h31) at (4,-4) [label=right:${u_{4r}^{ 4r }}$] {};
\vertex (h32) at (2,-4) [label=left:${u_{4r}^{ 1}}$] {};

\path 

(h1) edge (h)
(h1) edge (h11)
(h11) edge (h21)
(h11) edge (h22)
(h1) edge (h12)
(h12) edge (h31)
(h12) edge (h32);

\draw[dotted, thick] (-2,-2.5) -- (2,-2.5);
\draw[dotted, thick] (-3.5,-4) -- (-2.5,-4);
\draw[dotted, thick] (3.5,-4) -- (2.5,-4);

 \end{tikzpicture}
    \caption{Our tree gadget $T_{u}$ for each degree one forbidden vertex $u\in V_{\square}$}
    \label{fig:OA4}
\end{figure}

\proof We  give a parameterized reduction from   {\sc Offensive Alliance$^{\mbox{F}}$} which is 
 W[1]-hard when parameterized by the size of a vertex deletion set into trees of height at most 5. 
Let $I=(G, r, V_{\square})$ be an instance of {\sc Offensive Alliance$^{\mbox{F}}$}. 
Let $n=|V(G)|$. 
We construct an instance $I'=(G',r')$ of {\sc Offensive Alliance} the following way. We
set $r'=r$. Recall that each degree one
forbidden vertex is adjacent to another forbidden vertex and each  forbidden vertex of degree greater than one is adjacent to a degree one forbidden vertex. Let $u$ be a degree one forbidden vertex in $G$
and $u$ is adjacent to another forbidden vertex $v$.  
For each degree one forbidden vertex $u\in V_{\square}$, 
we introduce a tree 
$T_{u}$ rooted at $u$ of height 2 as shown in 
Figure \ref{fig:OA4}. The forbidden vertex $v$ has additional neighbours from the original 
graph $G$ which are not shown here. We define $G'$ as follows: 
$$ V(G') = V(G) \bigcup\limits_{u\in V_{\square}} \Big \{V(T_{u}) ~|~ \text{where } u  \text{ is a  degree one forbidden vertex in } G \Big\} 
$$
and
\begin{align*}
        E(G') = E(G) \bigcup\limits_{u'\in V_{\square}} {E(T_{u})}. 
\end{align*} 

We claim $I$ is a yes instance if and only if $I'$ is a yes instance. It is easy to see that 
if $R$ is an offensive alliance of size at most $r$ in $G$ then it is also an 
offensive alliance of size at most $r'=r$ in $G'$. 
\par To prove the reverse direction of the equivalence, suppose that $G'$ has an  offensive alliance $R'$  of size 
at most $r'=r$. We claim that no vertex from the set $V_{\square} \bigcup\limits_{u\in V_{\square}} V(T_{u})$  
is part of $R'$.
It is easy to see that if any vertex from the set $V_{\square} \bigcup\limits_{u\in V_{\square}} V(T_{u})$ is in $R'$ then the size of $R'$ exceeds $2r$. This implies that $R=R' \cap G$ is an offensive alliance such that $R\cap V_{\square}=\emptyset$ and $|R|\leq r$. This shows that $I$ is a yes instance.\qed\\

We have the following consequences. 
\begin{corollary}\label{corollary4}\rm
 The {\sc Exact Offensive Alliance} problem is W[1]-hard when parameterized by the size of a vertex deletion set into trees of height at most 7.
 \end{corollary}
 
\noindent  Clearly trees of height at most seven are trivially acyclic. 
 Moreover, it is easy to verify that such trees have 
 pathwidth \cite{Kloks94} and treedepth \cite{Sparsity} at most seven, which implies:
 
\begin{theorem}\rm
 The {\sc Offensive Alliance} and {\sc Exact Offensive Alliance} problems
 are W[1]-hard when parameterized by any of the following parameters:
 \begin{itemize}
     \item the feedback vertex set number,
     \item the treewidth and pathwidth of the input graph,
     \item the treedepth of the input graph.
 \end{itemize}
\end{theorem}

\section{Conclusions} 
 In this work we proved that 
 the {\sc Offensive Alliance} problem is W[1]-hard
 parameterized by a wide range of 
fairly restrictive structural parameters such as the feedback vertex set number, treewidth, pathwidth, and treedepth of the input graph  thus not FPT (unless FPT = W[1]).
 This is especially interesting because most ``subset problems'' that are FPT when parameterized by solution size turned out to be FPT for the parameter treewidth \cite{DomMichael}, and moreover {\sc Offensive Alliance} is easy on trees. 
  In the future it may be interesting to study if our ideas can be useful for different kinds of alliances from the literature such as powerful alliances and defensive alliances.
 It would be interesting to consider the parameterized complexity with respect to twin cover.    The parameterized complexity of  offensive and defensive alliance problems 
remain unsettled  when parameterized by other important 
structural graph parameters like clique-width and modular-width.

\bibliographystyle{abbrv}
\bibliography{bibliography}
\end{document}